\RequirePackage{fix-cm}
\documentclass[twocolumn,epjc3]{svjour3}  
\smartqed  
\RequirePackage{graphicx}
\RequirePackage[numbers,sort&compress]{natbib}
\journalname{Eur. Phys. J. C}
\begin{document}

\title{Excited state mass spectra of doubly heavy $\Xi$ baryons
}


\author{Zalak Shah\thanksref{e1}
        \and
        Ajay Kumar Rai\thanksref{e2}} 

\thankstext{e1}{e-mail: zalak.physics@gmail.com}
\thankstext{e2}{e-mail: raiajayk@gmail.com}

\institute{Department of Applied Physics, Sardar Vallabhbhai National Institute of Technology, Surat, Gujarat, India-395007 
}

\date{Received: date / Accepted: date}

\maketitle

\begin{abstract}
In this paper, the mass spectra are obtained for doubly heavy $\Xi$ baryons, namely, $\Xi_{cc}^{+}$, $\Xi_{cc}^{++}$, $\Xi_{bb}^{-}$, $\Xi_{bb}^{0}$, $\Xi_{bc}^{0}$ and $\Xi_{bc}^{+}$. These baryons are consist of two heavy quarks($cc$, $bb$ and $bc$) with a light($d$ or $u$) quark. The ground, radial and orbital states are calculated in framework of Hypercentral constituent quark model with coul- omb plus linear potential. Our outcomes are also compared with other predictions, thus, the average possible range of excited states masses of these $\Xi$ baryons can be determined. The study of the Regge trajectories are performed in (n, $M^{2}$) and (J, $M^{2}$) planes and their slopes and intercepts are also determined. Lastly, the ground state magnetic moments of these doubly heavy baryons are also calculated.
\end{abstract}

\section{Introduction}
Doubly heavy baryons have two families: $\Xi$ and $\Omega$. $\Omega$ has a light strange quark while $\Xi$ has up or down quark(s) with two heavy quarks(c and b). Our previous work \cite{zalak1} exhibited the mass spectra, magnetic moments and Regge trajectories of doubly heavy $\Omega$ baryons while in present paper, we established the $\Xi$ baryon family with six members. The only experimental evidence comes for $\Xi_{cc}^{+}$ by the SELEX experiment. They reported a ground state at 3520 MeV containing two charm quarks and a down quark \cite{olive,selex,selex1}. It's yet to be confirmed from the other experiments \cite{babar,belle,lhcb,focus}. Recently,  Hamiltonian model \cite{yoshida}, Regge phenomenology \cite{kwei}, Lattice QCD \cite{paula,brown,alex,mathur}, QCD sum rule \cite{aliev}, variational approach \cite{roberts} and many more {\cite{Roberts2008,Giannuzzi2009,valcarce,ebert,Gershtein,kar,Ltang,albertus,1,wang,patel} are providing new results in field of doubly heavy baryons. Many of them have only calculated the ground state masses while some of them have also shown the excited states. 

We have used the QCD inspired Hypercentral constituent quark model(hCQM) with coloumb plus linear potential. The first order correction is also taken into account to the potential and calculation has been performed by solving six dimensional hyper radial Schr$\ddot{o}$-dinger equation numerically \cite{zalak,zalak1,zalak3}. We have calculated the mass spectra of radial excited states upto 5S and orbital excited states for 1P-5P, 1D-4D and 1F-2F states. As per our knowledge, all the theoretical approaches have considered the $m_u$=$m_d$ so far but the light quark masses are different in our model. Thus, we have obtained the mass spectra with u \& d quarks combinations for these baryons. Obtained masses were used in formation of Regge trajectories in (n, $M^{2}$) and (J, $M^{2}$) planes. The determination of slope and intercept of Regge trajectories of these baryons are very important as it provides better understanding of the dynamics of strong interactions in production of charmed and bottom baryons at high energies.

The paper is organized as follows. We give brief introduction in sect.1 and explained our Hypercentral Constituent Quark Model in sect. 2.  We present our mass spectra results of all doubly heavy $\Xi$ baryons in sect. 3. Regge trajectories and magnetic moments are discussed in sect. 4. At last, our conclusion is in sect. 5.
\begin{table*}
\centering
\caption{\label{table:1} Ground state masses of $\Xi$ baryons are listed and compared. Our work shows the ground state masses of all baryons with both light quark combination as $\Xi_{cc}^{+}$/$\Xi_{cc}^{++}$, $\Xi_{bb}^{-}$/$\Xi_{bb}^{0}$ and $\Xi_{bc}^{0}$/$\Xi_{bc}^{+}$  (in GeV).}
\begin{tabular}{@{}lllllll}
\hline\noalign{\smallskip}    
Baryons& \multicolumn{1}{c}{$\Xi_{ccd}$/$\Xi_{ccu}$}&\multicolumn{3}{c}{$\Xi_{bbd}$/$\Xi_{bbu}$}&\multicolumn{1}{c}{$\Xi_{bcd}$/$\Xi_{bcu}$}\\
 $J^{P}$& $\frac{1}{2}^{+}$ &$\frac{3}{2}^{+}$ & $\frac{1}{2}^{+}$ &$\frac{3}{2}^{+}$& $\frac{1}{2}^{+}$ &$\frac{3}{2}^{+}$  \\
\hline
Our work&3.520/3.511&3.695/3.687&10.317/10.312&10.340/10.335&6.920/6.914&6.986/6.980\\
Exp.\cite{olive}&\multicolumn{1}{c}{3.519$\pm$0.009}&&-&--&-&-\\
 Ref.\cite{yoshida}&3.685&3.754&10.314&10.339&-&-\\
 Ref.\cite{kwei}&3.520&3.695&10.199&10.316&-&-\\
Ref.\cite{paula}&3.610(09)(12)&3.694(07)(11)&-&-&-&-\\
Ref.\cite{brown}&3.610&3.692&10.143&10.178&6.943&6.985\\
Ref.\cite{alex}&3.561(22)&3.642(26)&-&-&-&-\\
Ref.\cite{aliev}&3.720&-&9.960&-&6.720&-\\
Ref.\cite{roberts}&3.678&3.752&10.322&10.352&7.014&7.064\\
Ref.\cite{Roberts2008}&3.676&3.753&10.340&10.367&7.011&7.074\\
Ref.\cite{Giannuzzi2009}&3.547&3.719&10.185&10.216&6.904&6.936\\
Ref.\cite{valcarce}&3.579&3.656&10.189&10.218&-&-\\
Ref.\cite{ebert}&3.620&3.727&10.202&10.237&6.933&6.980\\
Ref.\cite{Gershtein}
&3.478&3.610&10.093&10.133&6.820&6.900\\
Ref.\cite{kar}&3.627&3.690&10.162&10.184&6.914\\
Ref.\cite{Ltang}&3.519&3.620&9.800&9.890&6.650&6.690\\
Ref.\cite{albertus}&3.612&3.706&10.197&10.136&6.919&6.986\\
Ref.\cite{1}&3.510&3.548&10.130&10.144&6.792&6.827\\
Ref.\cite{wang}&3.570&3.610&10.170&10.220&-&-\\
\noalign{\smallskip}\hline
\end{tabular}
\end{table*}
\section{The Model}
The methodology for the determination of excited masses follow the same pattern as in our previous work \cite{zalak1} and Refs. their-in. Therefore, we discuss model very briefly in present paper. Starting with the Jacobi co-ordinates of three quark baryons that are given in terms of mass($m_i$) and co ordinates($\vec{r_i}$) below \cite{Bijker}.  The quark masses are taken in calculations as $m_{u}$ = 0.338, $m_{d}$ = 0.350, $m_{c}$ = 1.275 and $m_{b}$ = 4.67 (in GeV). The co-ordinates $\vec{\rho}$ and $\vec{\lambda}$ are with the respective reduced masses are given by

\begin{equation}
m_{\rho}=\frac{2 m_{1} m_{2}}{m_{1}+ m_{2}}, 
\end{equation}
\begin{equation}
 m_{\lambda}=\frac{2 m_{3} (m_{1}^2 + m_{2}^2+m_1m_2)}{(m_1+m_2)(m_{1}+ m_{2}+ m_{3})}
\end{equation}
The  Hamiltonian of three body baryonic system in the hCQM is then expressed as
\begin{equation}
H=\frac{P_{x}^{2}}{2m} +V(x)
\end{equation}
The hyper radius $x= \sqrt{\rho^{2} + \lambda^{2}}$ is a collective co ordinate and therefore the hypercentral potential contains also the three-body effects. 
where, $m=\frac{2 m_{\rho} m_{\lambda}}{m_{\rho} + m_{\lambda}}$, is the reduced mass and $x$ is the six dimensional radial hyper central coordinate of the three body system. In present paper, the confining three-body potential is chosen within a string-like picture, where the quarks are connected by gluonic strings and the potential strings increases linearly with a collective radius $r_{3q}$ as mentioned in \cite{ginnani2015}. Accordingly the effective two body interactions can be written as
\begin{table*}
\centering
\caption{\label{table:2}Radial excited states of doubly heavy $\Xi$ baryons(in GeV). Column 4 and 5 show the masses with light quark d combination whereas Column 6 and 7 show the masses with light quark u. } 
\begin{tabular}{@{}lllllllllllllll}
\hline\noalign{\smallskip} Particle&State & $J^{P}$ & A &B&A &B &\cite{yoshida}&\cite{Roberts2008}&\cite{Giannuzzi2009}&\cite{valcarce}&\cite{ebert}&\cite{roberts}\\
\hline
\multicolumn{11}{c}{A$\rightarrow$without first order correction, B$\rightarrow$with first order correction }\\
\noalign{\smallskip}    
&2S	&&	3.912&3.925&3.905&3.920	&4.079&4.029&4.183	&3.976&3.910&4.030\\
&3S	&$\frac{1}{2}^{+}$	&4.212	&4.233&4.230	&4.159&4.206&&4.640&&4.154	\\
$\Xi_{ccd}$&4S	&	&4.473	&4.502&4.468&4.501		&	\\
and&5S	&	&4.711	&4.748	&4.708&4.748	\\
\noalign{\smallskip}
$\Xi_{ccu}$&2S&	&3.976	&3.988&3.970&3.983	&4.114&4.042&4.282&4.025&4.027&4.078	\\
&3S&$\frac{3}{2}^{+}$	&4.244	&4.264&4.238&4.261	&4.131&&4.719	\\
&4S	&	&	4.492	&4.520	&4.488&4.519	\\
&5S&	&	4.724	&4.759	&4.720&4.759	\\
\noalign{\smallskip}
&2S	&&	10.605	&	10.612&10.603&10.609	&10.571&10.576&10.751&10.482&10.441&10.551\\
&3S	&$\frac{1}{2}^{+}$	&	10.851	&	10.862&10.851&10.862	&10.612&&11.170	&&10.630\\
$\Xi_{bbd}$&4S	&	&	11.073	&	11.088&11.075&11.090	&&&&&10.812	\\
and&5S	&	&	11.278	&	11.297	&11.282&11.301	\\
\noalign{\smallskip}
$\Xi_{bbu}$&2S&	&	10.613	&	10.619&10.611&10.617	&10.592&10.578&10.770&10.501&10.482&10.574	\\
&3S&$\frac{3}{2}^{+}$	&	10.855&10.855&10.866	&	10.866	&10.593&&11.184	&&10.673\\
&4S	&	&	11.075	&	11.090&11.077&11.092	&&&&&10.856	\\
&5S&	&	11.280	&	11.298	&11.284&11.302	\\
\noalign{\smallskip}
&2S	&&7.236		&7.244&7.231&7.240		&&&7.478&&&7.321\\
&3S	&$\frac{1}{2}^{+}$	&7.495	&	7.509&7.492&7.507	&&&7.904&\\
$\Xi_{bcd}$&4S	&	&7.727		&7.746&7.726&7.744	&&&&&	\\
and&5S	&	&	7.940	&7.963		&7.940&7.964	\\
\noalign{\smallskip}
$\Xi_{bcu}$&2S&	&7.260	&7.267	&7.256&7.263&		&&7.495&&&7.353	\\
&3S&$\frac{3}{2}^{+}$	&7.507	&7.521&7.505&7.518		&&&7.917&\\
&4S	&	&7.734	&7.752&7.733&7.752	\\
&5S&	&7.944	&7.968&7.945&7.969	\\
\noalign{\smallskip}\hline
\end{tabular}
\end{table*}
\begin{equation}
\sum_{i<j}V(r_{ij})=V(x)+. . . .
\end{equation}
In the hypercentral approximation, the potential is only depends on hyper radius(x). More details can be seen in references \cite{ginnani2015,M. Ferraris}. We consider a reduced hypercentral radial function, $\phi_{\gamma}(x) = x^{\frac{5}{2}}\Psi_{ \gamma}(x)$ where $\Psi_{\gamma}$(x) is the hypercentral wave function and $\gamma$ is the grand angular quantum number. Thus, six dimensional hyperradial Schrodinger equation reduces to,
\begin{equation}\label{eq:6}
\left[\frac{-1}{2m}\frac{d^{2}}{d x^{2}} + \frac{\frac{15}{4}+ \gamma(\gamma+4)}{2mx^{2}}+ V(x)\right]\phi_{ \gamma}(x)= E\phi_{\gamma}(x)
\end{equation}

For the present study we consider the hypercentral potential $V(x)$ as the color coulomb plus linear potential with first order correction \cite{koma,11} is given below.
\begin{equation}\label{eq:7}
V(x) =  V^{0}(x) + \left(\frac{1}{m_{\rho}}+ \frac{1}{m_{\lambda}}\right) V^{(1)}(x)+V_{SD}(x) 
\end{equation}
\begin{equation}
V^{(0)}(x)= \frac{\tau}{x}+ \beta x ~~~~~~~~~~~~~\&~~~~~~~~~ 
V^{(1)}(x)= - C_{F}C_{A} \frac{\alpha_{s}^{2}}{4 x^{2}}
\end{equation}
\begin{eqnarray}
V_{SD}(x)= V_{SS}(x)(\vec{S_{\rho}}.\vec{S_\lambda})
+ V_{\gamma S}(x) (\vec{\gamma} \cdot \vec{S})&&  \nonumber \\ + V_{T} (x)
\left[ S^2-\frac{3(\vec{S }\cdot \vec{x})(\vec{S} \cdot \vec{x})}{x^{2}} \right]
\end{eqnarray}
Here, $\tau$ is the hyper-Coulomb strength corresponds to the strong running coupling constant $\alpha_{s}$. $\beta$ is the string tension of the confinement part of potential. $C_{F}$ and $C_{A}$ are the Casimir charges of the fundamental and adjoint representation with values $\frac{2}{3}$ and 3. The spin-dependent part, $V_{SD}(x)$  contains three types of the interaction terms \cite{12}: the spin-spin term $V_{SS} (x)$, the spin-orbit term $V_{\gamma S}(x)$ and tensor term $V_{T}(x)$. The detail of the terms are given in \cite{zalak}. 

\section{Mass Spectroscopy: $\Xi_{cc}$, $\Xi_{bb}$ \& $\Xi_{bc}$ }

\begin{table*}
\caption{\label{table:3} P and D state masses of $\Xi_{cc}$ baryon(in GeV).}
\begin{tabular}{@{}lllllllllllll}
\hline\noalign{\smallskip}    
State& A &B& A &B  &\cite{yoshida}&\cite{Roberts2008}&\cite{valcarce}&\cite{ebert}&\cite{kwei}&\cite{Gershtein}&\cite{roberts}&\cite{paula}\\
&\multicolumn{2}{l}{$\Xi_{cc}^{+}$}&\multicolumn{2}{l}{$\Xi_{cc}^{++}$}&&&&&&\\
\hline
\multicolumn{11}{l}{A$\rightarrow$ without correction and B$\rightarrow$ with correction}\\
\noalign{\smallskip} 
$(1^2P_{1/2})$&	3.853	&	3.865&	3.846	&	3.861	&3.947	&3.910&3.880&3.838&&&4.073&3.892(47)(48)	\\
$(1^2P_{3/2})$&	3.834	&	3.847&	3.828&	3.842&3.949	&3.921&&3.959&3.786&3.834&4.079&3.989(58)(58)	\\
$(1^4P_{1/2})$&	3.862	&	3.875	&3.856	&3.871	\\
$(1^4P_{3/2})$&	3.843	&	3.856	&3.837	&3.851	\\
$(1^4P_{5/2})$&	3.818	&	3.890&3.817&3.888	&4.163	&4.092	&&4.155&3.949&4.047&4.089\\
\noalign{\smallskip} 
$(2^2P_{1/2})$&	4.138	&	4.161&4.134&4.140&4.135	&4.074&4.018&4.085	\\
$(2^2P_{3/2})$&	4.121	&	4.144&4.118&4.140	&4.137	&4.078&4.197	\\
$(2^4P_{1/2})$&	4.146	&	4.169	&4.143&4.167	&	\\
$(2^4P_{3/2})$&	4.130	&	4.152	&4.126&4.149	&	\\
$(2^4P_{5/2})$&	4.108	&	4.183&4.104&4.181&4.488	\\
\noalign{\smallskip} 
$(3^2P_{1/2})$&	4.395	&	4.426&4.393&4.409	&4.149	&	\\
$(3^2P_{3/2})$&	4.381	&	4.411&4.379&4.409	&4.159	&	\\
$(3^4P_{1/2})$&	4.402	&	4.433	&4.400	&4.432	\\
$(3^4P_{3/2})$&	4.388	&	4.419	&4.386	&4.417	\\
$(3^4P_{5/2})$&	4.369	&	4.399&4.412&4.396	&4.534		\\
\noalign{\smallskip} 
$(4^2P_{1/2})$&	4.633	&	4.671	&4.633	&4.671	\\
$(4^2P_{3/2})$&	4.620	&	4.658	&4.620	&4.657	\\
$(4^4P_{1/2})$&	4.640	&	4.678	&4.639	&4.678	\\
$(4^4P_{3/2})$&	4.627	&	4.664	&4.626	&4.664	\\
$(4^4P_{5/2})$&	4.610	&	4.646	&4.609	&4.646	\\
\noalign{\smallskip} 
$(5^2P_{1/2})$&	4.857	&	4.901	&4.858	&4.902	\\
$(5^2P_{3/2})$&	4.845	&	4.889	&4.846	&4.889	\\
$(5^4P_{1/2})$&	4.863	&	4.908	&4.864	&4.909	\\
$(5^4P_{3/2})$&	4.851	&	4.895	&4.852	&4.896	\\
$(5^4P_{5/2})$&	4.835	&	4.878	&4.835	&4.879	\\
\noalign{\smallskip} 
$(1^4D_{1/2})$&	4.053	&	4.077	&4.043&4.071	\\
$(1^2D_{3/2})$	&	4.026	&	4.049	&4.019	&4.044	\\
$(1^4D_{3/2})$&	4.035	&	4.058&4.027&4.053&&	&&&	\\
$(1^2D_{5/2})$&	4.002	&4.024&3.998&4.019&4.043	&4.115	&4.047	&&4.391&4.034&4.050&4.388	\\
$(1^4D_{5/2})$&	4.011	&	4.033&4.006&4.029&4.027&4.052&&&&	\\
$(1^4D_{7/2})$	&	3.982	&	4.002&3.979	&3.998&4.097	&&&4.187&4.089&4.393\\
\noalign{\smallskip} 
$(2^4D_{1/2})$&	4.311	&	4.345	&4.311	&4.342	\\
$(2^2D_{3/2})$	&	4.289	&	4.321	&4.287	&4.318	\\
$(2^4D_{3/2})$&	4.296	&	4.329	&4.295	&4.326	\\
$(2^2D_{5/2})$&	4.270	&	4.299&4.267&4.297	&4.164	&4.091	\\
$(2^4D_{5/2})$&	4.277	&	4.307	&4.275	&4.305	\\
$(2^4D_{7/2})$	&	4.253	&	4.280&4.249	&4.278	&4.394	\\
\noalign{\smallskip} 
$(3^4D_{1/2})$&	4.554	&	4.592	&4.553	&4.592	\\
$(3^2D_{3/2})$	&	4.534	&	4.571	&4.532	&4.570	\\
$(3^4D_{3/2})$&	4.541	&	4.578	&4.539	&4.578	\\
$(3^2D_{5/2})$&	4.516	&	4.552&4.514&4.551	&4.348	&	\\
$(3^4D_{5/2})$&	4.523	&	4.559	&4.521	&4.558	\\
$(3^4D_{7/2})$	&	4.500	&	4.535	&4.498	&4.534	\\
\noalign{\smallskip} 
$(4^4D_{1/2})$&	4.780	&	4.825	&4.781	&4.826	\\
$(4^2D_{3/2})$	&	4.762	&	4.806	&4.762	&4.806	\\
$(4^4D_{3/2})$&	4.768	&	4.812	&4.768	&4.813	\\
$(4^2D_{5/2})$&	4.745	&	4.788	&4.745	&4.788	\\
$(4^4D_{5/2})$&	4.751	&	4.795	&4.751	&4.795&&&	\\
$(4^4D_{7/2})$	&	4.731	&	4.772	&4.730	&4.772&&&	\\
\noalign{\smallskip}\hline
\end{tabular}
\end{table*}
\begin{table*}
\caption{\label{table:4}P and D state masses of $\Xi_{bb}$ (in GeV).}
\begin{tabular}{@{}llllllllllll}
\hline\noalign{\smallskip}    
State&A &B&A&B  &\cite{yoshida}&\cite{Roberts2008}&\cite{valcarce}&\cite{ebert}&\cite{kwei}&\cite{roberts}&Others\\
&\multicolumn{2}{l}{$\Xi_{bb}^{-}$}&\multicolumn{2}{l}{$\Xi_{bb}^{0}$}&&&&&&\\
\hline
\multicolumn{7}{l}{A$\rightarrow$ without correction and B$\rightarrow$ with correction}\\
\noalign{\smallskip}    
$(1^2P_{1/2})$	&	10.507	&	10.514&10.504&10.511	&10.476	&10.493&10.406	&10.368&&10.691\\
$(1^2P_{3/2})$	&	10.502	&	10.509&10.499&10.506	&10.476	&10.495&&10.408&10.474&10.692&10.390\cite{wang}	\\
$(1^4P_{1/2})$	&	10.510	&	10.517	&10.506&10.514		\\
$(1^4P_{3/2})$	&	10.505	&	10.512&10.501&10.509	&&&&	&&&10.430\cite{aliev}	\\
$(1^4P_{5/2})$	&	10.514	&	10.521&10.512&10.518	&10.759	&&&&10.588&10.695	\\
\noalign{\smallskip}    
$(2^2P_{1/2})$	&	10.758	&	10.770&10.757&10.770	&	10.703&10.710&10.612&10.563	\\
$(2^2P_{3/2})$	&	10.754	&	10.766&10.753&10.765	&10.704&10.713&&10.607		\\
$(2^4P_{1/2})$	&	10.760	&	10.772	&10.760&10.772		\\
$(2^4P_{3/2})$	&	10.756	&	10.768	&10.755	&10.767	\\
$(2^4P_{5/2})$	&	10.751	&	10.763&10.763&10.776	&10.973	&10.713	\\
\noalign{\smallskip}    
$(3^2P_{1/2})$	&	10.985	&	11.001&10.986&11.002	&10.740	&&&10.744	\\
$(3^2P_{3/2})$	&	10.981	&	10.997&10.982&10.998	&10.742&&&10.788	\\
$(3^4P_{1/2})$	&	10.987	&	11.003	&10.988&11.004		\\
$(3^4P_{3/2})$	&	10.983	&	10.999	&10.984&11.000		\\
$(3^4P_{5/2})$	&	10.978	&	10.994&10.991&11.007	&11.004	\\
\noalign{\smallskip}    
$(4^2P_{1/2})$	&	11.194	&	11.214&11.197&11.217	&&&&10.900	\\
$(4^2P_{1/2})$	&	11.191	&	11.210	&11.194&11.213		\\
$(4^2P_{3/2})$	&	11.196	&	11.216	&11.199&11.219		\\
$(4^4P_{3/2})$	&	11.193	&	11.212	&11.796&11.215		\\
$(4^4P_{5/2})$	&	11.188	&	11.208	&	11.202&11.222	\\
\noalign{\smallskip}    
$(5^2P_{1/2})$	&	11.390	&	11.413	&11.395&11.418		\\
$(5^2P_{3/2})$	&	11.387	&	11.410	&11.392&11.415		\\
$(5^4P_{1/2})$	&	11.392	&	11.415	&11.397&11.420		\\
$(5^4P_{3/2})$	&	11.389	&	11.412	&11.394&11.417		\\
$(5^4P_{5/2})$	&	11.385	&	11.407	&11.399&11.423		\\
\noalign{\smallskip}    
$(1^4D_{1/2})$		&	10.665	&	10.677	&10.663&10.675		\\
$(1^2D_{3/2})$		&	10.658	&	10.670	&10.656&10.668		\\
$(1^4D_{3/2})$		&	10.660	&	10.672&10.659&10.670	&&&&&&11.011		\\
$(1^2D_{5/2})$		&	10.652	&	10.663&10.650&10.661	&10.592&10.676&&&10.742&11.002		\\
$(1^4D_{5/2})$		&	10.654	&	10.666	&10.652&10.664		\\
$(1^4D_{7/2})$		&	10.647	&	10.658&10.644&10.656	&&10.608&&&10.853&11.011		\\
\noalign{\smallskip}    
$(2^4D_{1/2})$	&	10.897	&	10.913	&10.897&10.913		\\
$(2^2D_{3/2})$	&	10.891	&	10.907	&10.891&10.907		\\
$(2^4D_{3/2})$	&	10.893	&	10.909	&10.893&10.909		\\
$(2^4D_{3/2})$	&	10.886	&	10.901&10.886&10.901	&&10.712		\\
$(2^2D_{5/2})$	&	10.888	&	10.903&10.888&10.903	&10.613		\\
$(2^2D_{5/2})$	&	10.881	&	10.896&10.881&10.896	&&11.057		\\
\noalign{\smallskip}    
$(3^4D_{1/2})$	&	11.109	&	11.130	&11.113&11.133		\\
$(3^2D_{3/2})$	&	11.105	&	11.125	&11.108&11.127		\\
$(3^4D_{3/2})$	&	11.107	&	11.126	&11.110&11.129		\\
$(3^4D_{3/2})$	&	11.101	&	11.120	&11.102&11.122		\\
$(3^2D_{5/2})$	&	11.103	&	11.122&11.105&11.124	&10.809		\\
$(3^2D_{5/2})$	&	11.097	&	11.116	&11.099&11.118		\\
\noalign{\smallskip}    
$(4^4D_{1/2})$	&	11.310	&	11.333	&11.314&11.337		\\
$(4^2D_{3/2})$	&	11.306	&	11.328	&11.310&11.332		\\
$(4^4D_{3/2})$	&	11.307	&	11.330	&11.311&11.334		\\
$(4^4D_{3/2})$	&	11.302	&	11.324	&11.305&11.328		\\
$(4^2D_{5/2})$	&	11.303	&	11.325	&11.307&11.330		\\
$(4^2D_{5/2})$	&	11.298	&	11.320	&11.302&11.324		\\
\noalign{\smallskip}\hline
\end{tabular}
\end{table*}

\begin{table*}
\centering
\caption{\label{table:5}P and D state masses of $\Xi_{bc}$ (in GeV).}
\begin{tabular}{@{}llllllllllll}
\hline\noalign{\smallskip}    
State&A &B&A &B&\cite{roberts}&State&A &B&A &B&\cite{roberts}  \\
&\multicolumn{2}{l}{$\Xi_{bc}^{0}$}&\multicolumn{2}{l}{$\Xi_{bc}^{+}$}&&&\multicolumn{2}{l}{$\Xi_{bc}^{0}$}&\multicolumn{2}{l}{$\Xi_{bc}^{+}$}&\\
\hline
\multicolumn{7}{l}{A$\rightarrow$ without correction and B$\rightarrow$ with correction}\\
\noalign{\smallskip}
$(1^2P_{1/2})$	&7.151&7.160&7.146&7.156&7.390&$(1^4D_{1/2})$&7.322	&	7.336&7.318&7.334\\
$(1^2P_{3/2})$	&7.140&7.149&7.135&7.144&7.394&$(1^2D_{3/2})$&7.307	&	7.321&7.303&7.318	\\
$(1^4P_{1/2})$	&	7.157&7.166&7.152&7.161&7.399	&$(1^4D_{3/2})$&7.312&7.326&7.706&7.308&7.324	\\
$(1^4P_{3/2})$	&7.146&7.155&7.141&7.150&&$(1^2D_{5/2})$	&7.294&7.308&7.290&7.304	\\
$(1^4P_{5/2})$	&7.131&7.175&7.126&7.171&&$(1^4D_{5/2})$	&7.299&7.313&7.702&7.295&7.309	\\
&&&&&&$(1^4D_{7/2})$&7.293&7.296&7.708&7.278&7.292\\
\noalign{\smallskip}
$(2^2P_{1/2})$	&7.410&7.425&7.407&7.422&&$(2^4D_{1/2})$&7.559	&7.425&7.558&7.579&	7.579		\\
$(2^2P_{3/2})$	&7.401	&7.415&7.397&7.412&&$(2^2D_{3/2})$&7.547&7.567&7.545&7.565
		\\
$(2^4P_{1/2})$	&	7.415	&	7.430&7.411&7.426&&$(2^4D_{3/2})$&7.551&7.571&7.549&7.570\\
$(2^4P_{3/2})$	&7.405	&	7.420&7.402&7.417&&$(2^2D_{5/2})$&7.536&7.555&7.534&7.553&7.538			\\
$(2^4P_{5/2})$	&7.393	&	7.408&7.419&7.434&&$(2^4D_{5/2})$&7.540&7.559&7.538&7.558\\
&&&&&&$(2^4D_{7/2})$&7.526&7.545&7.523&7.544\\
\noalign{\smallskip}
$(3^2P_{1/2})$	&7.643	&7.664&7.642&7.662&&$(3^4D_{1/2})$&7.779&7.804&7.777&7.804\\
$(3^2P_{3/2})$	&	7.635	&	7.655&7.634&7.654&&$(3^2D_{3/2})$&7.768&7.792&7.7668&7.792\\
$(3^4P_{1/2})$	&	7.647	&	7.668&7.646&7.666&&$(3^4D_{3/2})$&7.772&7.782&7.770&7.796\\
$(3^4P_{3/2})$	&7.639	&	7.659&7.638&7.658&&$(3^2D_{5/2})$&7.758&7.786&7.757&7.781\\
$(3^4P_{5/2})$	&	7.629	&	7.648&7.653&7.673&&$(3^4D_{5/2})$&7.762&7.772&7.761&7.785\\
&&&&&&$(3^4D_{7/2})$&7.774&7.772&7.749&7.772\\
\noalign{\smallskip}
$(4^2P_{1/2})$	&7.859	&	7.884&7.859&8.015&&$(4^4D_{1/2})$&7.985&		7.801&7.859&7.884&7.797\\
$(4^2P_{1/2})$	&7.852	&	7.876&7.852&7.877&&$(4^2D_{3/2})$&7.975	&	8.002&7.976&8.004\\
$(4^2P_{3/2})$	&7.863	&	7.888&7.863&7.888&&$(4^4D_{3/2})$&7.987	&8.006&7.979&8.008\\
$(4^4P_{3/2})$	&	7.856&	7.880&7.856&7.880&&$(4^2D_{5/2})$&7.969	&7.993&7.996&7.994	\\
$(4^4P_{5/2})$	&	7.846	&	7.870&7.869&7.895&&$(4^4D_{5/2})$&7.958	&7.996&7.970&7.998\\
&&&&&&$(4^4D_{7/2})$&	&7.985&7.958&7.986\\
\noalign{\smallskip}
$(5^2P_{1/2})$	&	8.062	&	8.091&8.064&8.092
	\\
$(5^2P_{3/2})$	&8.05528	&	8.084&8.057&8.085
		\\
$(5^4P_{1/2})$	&	8.0652&	8.094&8.067&8.096
	\\
$(5^4P_{3/2})$	&	8.059	&	8.087&8.060&8.088
	\\
$(5^4P_{5/2})$	&8.050	&	8.078&8.073&8.079
			\\
\noalign{\smallskip}\hline
\end{tabular}
\end{table*}

\begin{table*}
\centering
\caption{\label{table:6}F state masses of $\Xi_{cc}$,$\Xi_{bb}$ and $\Xi_{bc}$ baryons(in GeV).}
\begin{tabular}{@{}lllllllllllllllll}
\hline\noalign{\smallskip}    
&\multicolumn{2}{c}{$\Xi_{ccd}$}&\multicolumn{2}{c}{$\Xi_{ccu}$}&&\multicolumn{2}{c}{$\Xi_{bbd}$}&\multicolumn{2}{c}{$\Xi_{bbu}$}&&\multicolumn{2}{c}{$\Xi_{bcd}$}&\multicolumn{2}{c}{$\Xi_{bcu}$}\\
State& A &B&A&B &\cite{kwei}& A &B&A&B&\cite{kwei}  & A &B&A&B\\
\hline
$(1^4F_{3/2})$&	4.216	&	4.247&4.213&4.242	&		&	10.805	&	10.820&10.804&10.819	&&	7.468	&	7.487&7.466&7.485\\
$(1^2F_{5/2})$	&	4.185	&4.215&4.182&4.210		&		&	10.797	&	10.812&10.796&10.811&	&7.451	&	7.469&7.448&7.467\\
$(1^4F_{5/2})$	&	4.158	&4.186	&4.190&4.219		&		&10.790	&	10.804&10.798&10.813&	&7.456	&	7.474&7.453&7.472\\
$(1^2F_{7/2})$		&	4.166	&4.194&4.163&4.191		&		&10.799	&	10.814&10.791&10.806&&7.440	&	7.458&7.437&7.455\\
$(1^4F_{7/2})$		&	4.194	&4.225&4.154&4.182	&4.267		&10.792	&	10.806&10.789&10.803&11.004&7.435	&	7.453&7.432&7.450\\
$(1^4F_{9/2})$		&	4.133	&4.159&4.129&4.156		&4.413		&	10.784	&	10.797&10.783&10.797&11.112	&7.421	&	7.439&7.418&7.436\\
\hline
$(2^4F_{3/2})$	&	4.462	&4.494&4.460&4.497		&		&	11.023	&	11.022&11.024&11.043	&&7.692	&	7.715&7.691&7.715\\
$(2^2F_{5/2})$		&	4.435	&4.468&4.433&4.468		&		&	11.016	&	11.035&11.018&11.036&	&7.677	&	7.700&7.676&7.699\\
$(2^4F_{5/2})$		&	4.442	&4.475&4.440&4.476		&		&	11.018	&	11.036&11.019&11.038	&&7.681	&	7.704&7.680&7.703\\
$(2^2F_{7/2})$		&	4.410	&4.445	&4.415&4.450	&		&	11.010	&	11.028&11.013&11.031&	&7.667	&	7.690&7.666&7.689\\
$(2^4F_{7/2})$		&	4.417	&4.452&4.408&4.443		&		&	11.012	&	11.030&11.011&11.029&	&7.663	&	7.686&7.662&7.685\\
$(2^4F_{9/2})$		&	4.388	&4.424&4.386&4.420		&		&	11.005	&	11.022&11.006&11.023&	&	7.651	&	7.674&7.650&7.672\\

\noalign{\smallskip}\hline
\end{tabular}
\end{table*}
We begin with calculating the ground state masses of doubly heavy $\Xi_{cc}$, $\Xi_{bb}$ \& $\Xi_{bc}$ baryons\footnote{We compare our ccd, bbd and cbd baryon combination masses with others for whole mass spectra discussion}. The masses are computed for both parities $\frac{1}{2}^{+}$ and $\frac{3}{2}^{+}$ mentioned in Table~\ref{table:1}. As we know, the ground state of $\Xi_{cc}^{+}$ is found experimentally as  $\Xi_{cc}(3520)^{+}$; but its $J^{P}$ value is still undefined. Our prediction suggests that it would be $J^{P}= \frac{1}{2}^+$; similar suggestion given by refs. \cite{kwei,Ltang,1}. The other ground state, with $J^{P}= \frac{3}{2}^+$ is found as 3.695 GeV by us. The value is nearer to other predictions \cite{kwei,kar,albertus} and lattice \cite{paula,brown} calculations. In case of $\Xi_{bb}$ baryon, our ground state outcomes(both parities) are matched(with \cite{yoshida,roberts}) very well. Our predicted ground states of $\Xi_{bc}$ are very close to refs. \cite{ebert,albertus,brown}. We have also calculated the ground state spectra of ccu, bbu and bcu baryons. They are close to (8MeV, 5MeV and 6MeV difference, respectively) the results of d quark combinations.

Moving towards the excited states, the radial excited states are calculated from 2S-5S for $J^{P}$=$\frac{1}{2}^{+}$ and $\frac{3}{2}^{+}$. These radial excited states of $\Xi_{cc}$, $\Xi_{bb}$ \& $\Xi_{bc}$ are mentioned in Table~\ref{table:2}. P, D and F states with their different isospin splittings are computed and the excited state masses from 1P-5P, 1D-4D and 1F-2F are shown in Tables ~\ref{table:3},~\ref{table:4},~\ref{table:5} and ~\ref{table:6}. One can observe that B masses are few MeV higher than A masses in each case of every baryonic systems. The results of different theoretical approaches for all the systems are also compared in the respective tables. Notice that, the heavy quarks combination(cc,bb and bc) with light quarks(u and d) for all three baryon have represented individualy in Table ~\ref{table:1}-~\ref{table:6} .   

The excited states of doubly heavy $\Xi$ family are unknown experimentally. Starting from radial excited states mentioned in Table ~\ref{table:2}, we have compared our results with refs. \cite{yoshida,Roberts2008,roberts,valcarce,ebert,Giannuzzi2009}. For 2S state of the system $\Xi_{cc}$ with $J^{P}$ values $\frac{1}{2}^{+}(\frac{3}{2}^{+})$ have lowest prediction is 3.910 (4.027) (by \cite{ebert}) and highest is 4.183(4.282) (by\cite{Giannuzzi2009}). Specifically, our 2S predictions are close to \cite{ebert}.In similar way, while analyzing 2S state of $\Xi_{bb} (\frac{1}{2}^{+})$ and $\Xi_{bb} (\frac{3}{2}^{+})$ baryon, the lowest to highest range of masses are found to be (10441-10751)MeV and (10482-10770) MeV, respectively whereas our model sugeested masses are near to ref. \cite{Roberts2008} among all. The next baryon is $\Xi_{bc}$ and the 2S state with both isospins show more than 100 MeV difference with others(\cite{roberts, Giannuzzi2009}. Though Refs. \cite{yoshida,Giannuzzi2009,ebert} have computed 3S state(with $\frac{1}{2}^{+}$ and $\frac{3}{2}^{+}$) for $\Xi_{cc}$ and $\Xi_{bb}$ baryons and only \cite{Giannuzzi2009} has computed for $\Xi_{bc}$ baryon, but their obtained mass difference with us is large(except \cite{yoshida} for $\Xi_{cc}$). 

In case of $\Xi_{cc}$, our 1P state $J^{P}=\frac{1}{2}^{-}$ show difference of 57 MeV(with \cite{Roberts2008}), 27 MeV (with \cite{valcarce}), 12 MeV(with \cite{ebert}) and 39 MeV(with \cite{paula}), while $J^{P}=\frac{3}{2}^{-}$ shows 87 MeV(with \cite{Roberts2008}) and 0 MeV(with \cite{Gershtein}). 

For $\Xi_{bb}$, our 1P state $J^{P}=\frac{1}{2}^{-}$ and $J^{P}=\frac{3}{2}^{-}$ are only 14 and 7 MeV higher than ref. \cite{Roberts2008} whereas Ref. \cite{yoshida} masses for $J^{P}=\frac{1}{2}^{-}$ and $J^{P}=\frac{3}{2}^{-}$  are 31, 26  MeV lower than our prediction. Our 2P state is few MeV higher than \cite{yoshida,Roberts2008}. Our 1D-2D states have $\approx$35 MeV and $\approx$178 MeV difference with \cite{Roberts2008}. The P and D states of $\Xi_{bc}^{0}$ baryons are given in table~\ref{table:5} and it follow the same description mentioned in \cite{zalak1} and refs. therein. We have compare our results with recent paper \cite{kwei} for 1P and 1D. Their values are higher than us. The rest spectra (2P-5P and 2D-4D) is  performed by us for completeness.\\ 
 
F state masses for all three doubly heavy baryons are given in table~\ref{table:6}. Apart from us, Ref. \cite{kwei} has also calculated the 1F state of $\Xi_{cc}$ and $\Xi_{bb}$ for $J^{P}$= $\frac{7}{2}^{-}$ and $\frac{9}{2}^{-}$. For $\Xi_{cc}$, their masses are 73 and 280 MeV higher while for other system 212 and 328 MeV higher than our predictions. We did not find any other F state calculations for $\Xi_{bc}$ systems. 
 
\section{Regge Trajectories and Magnetic Moments}
\begin{table}
\centering
\caption{\label{table:7}Fitted slopes and intercepts of the regge trajectories in (n, $M^{2}$) plane.} 
\begin{tabular}{lllll}
\hline\noalign{\smallskip}            
Baryon&$J^{P}$ &State& $\beta$&$\beta_{0}$\\
\hline  
 $\Xi_{cc}$&$\frac{1}{2}^{+}$&S&0.409$\pm$0.0138&-5.181$\pm$0.247\\
&$\frac{1}{2}^{-}$&P&0.457$\pm$0.004 &-6.819$\pm$ 0.073\\
&$\frac{3}{2}^{-}$&P&0.456$\pm$0.003
&-6.729$\pm$0.075\\
&$\frac{5}{2}^{+}$&D&0.462$\pm$0.003&-7.408$\pm$0.058\\
\hline  
$\Xi_{bb}$&$\frac{1}{2}^{+}$&S&0.193
$\pm$0.006&-20.627$\pm$0.721\\
&$\frac{1}{2}^{-}$&P&0.206$\pm$0.004 &-22.914$\pm$0.540\\
&$\frac{3}{2}^{-}$&P&0.207
$\pm$0.045&-22.850$\pm$0.054\\
&$\frac{5}{2}^{+}$&D&0.210$\pm$0.004&-23.900$\pm$0.461\\
\hline  
$\Xi_{bc}$&$\frac{1}{2}^{+}$&S&0.265
$\pm$0.009&-12.777$\pm$0.501\\
&$\frac{1}{2}^{-}$&P&0.289$\pm$0.005 &-14.832$\pm$0.298\\
&$\frac{3}{2}^{-}$&P&0.288
$\pm$0.005&-14.738$\pm$0.305\\
&$\frac{5}{2}^{+}$&D&0.291$\pm$0.004&-15.522$\pm$0.215\\
\noalign{\smallskip}\hline 
\end{tabular}
\end{table}
As discussed in section 3, we have calculated the 1S-5S, 1P-5P, 1D-4D and 1F-2F state masses for all doubly heavy $\Xi$ baryons. The obtained masses were very much useful in constructing Regge trajectories in (n, $M^{2}$) and (J, $M^{2}$) planes. Where, n is principal quantum number and J is a total spin. The Regge trajectories are presented in Figs.~\ref{fig:0}-~\ref{fig:4} for $\Xi_{cc}(ccd)$, $\Xi_{bb}(bbd)$ and $\Xi_{bc}(bcd)$ baryons. Similar trajectories can also be plotted for the rest of the baryons. Straight lines were obtained by the linear fitting in all figures. The ground and radial excited S states ($J^{P}=\frac{1}{2}^{+}$) and the orbital excited P ($J^{P}= \frac{1}{2}^{-}$), D ($J^{P}= \frac{5}{2}^{+}$) and F ($J^{P}= \frac{7}{2}^{-}$) states are plotted in Fig~\ref{fig:0}-~\ref{fig:2} from bottom to top. We use,
\begin{figure}
\centering
\resizebox{0.55\textwidth}{!}{%
  \includegraphics{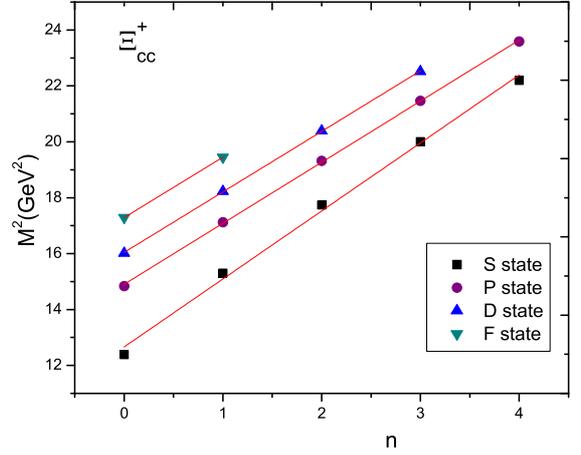}
}
\caption{Regge Trajectory ($M^{2}$ $\rightarrow$ n) for $\Xi_{cc}$ baryon.}
\label{fig:0}       
\end{figure}
\begin{equation}
n=\beta M^2+ \beta_{0}
\end{equation}
\noindent Where, $\beta$ and $\beta_{0}$ are slope and intercept, respectively.  The fitted slopes and intercepts are given in Table~\ref{table:7}}. 
\noindent We use natural parity $J^{P}=\frac{1}{2}^{+}$, $J^{P}=\frac{3}{2}^{-}$, $J^{P}=\frac{5}{2}^{+}$, $J^{P}=\frac{7}{2}^{-}$ and unnatural ($J^{P}=\frac{3}{2}^{+}$, $J^{P}=\frac{5}{2}^{-}$, $J^{P}=\frac{7}{2}^{+}$, $J^{P}=\frac{9}{2}^{-}$) parity masses and plotted graphs for $\Xi_{cc}$ and $\Xi_{bb}$ baryon states[See Fig. ~\ref{fig:3} and  ~\ref{fig:4}]. For that we use,
\begin{figure}
\centering

\resizebox{0.55\textwidth}{!}{%
  \includegraphics{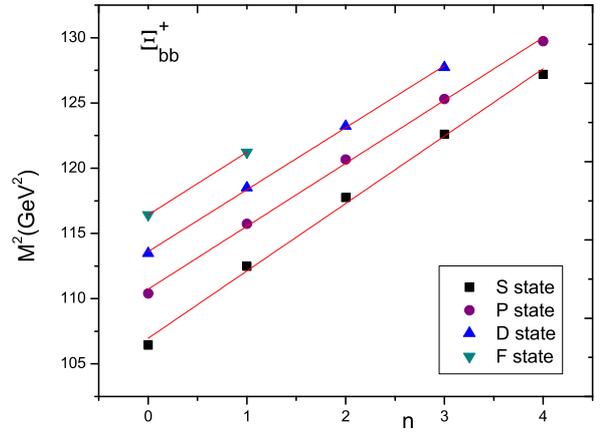}
}
\caption{Regge Trajectory ($M^{2}$ $\rightarrow$ n) for $\Xi_{bb}$ baryon.}
\label{fig:1}       
\end{figure}

\begin{figure}
\centering

\resizebox{0.55\textwidth}{!}{%
  \includegraphics{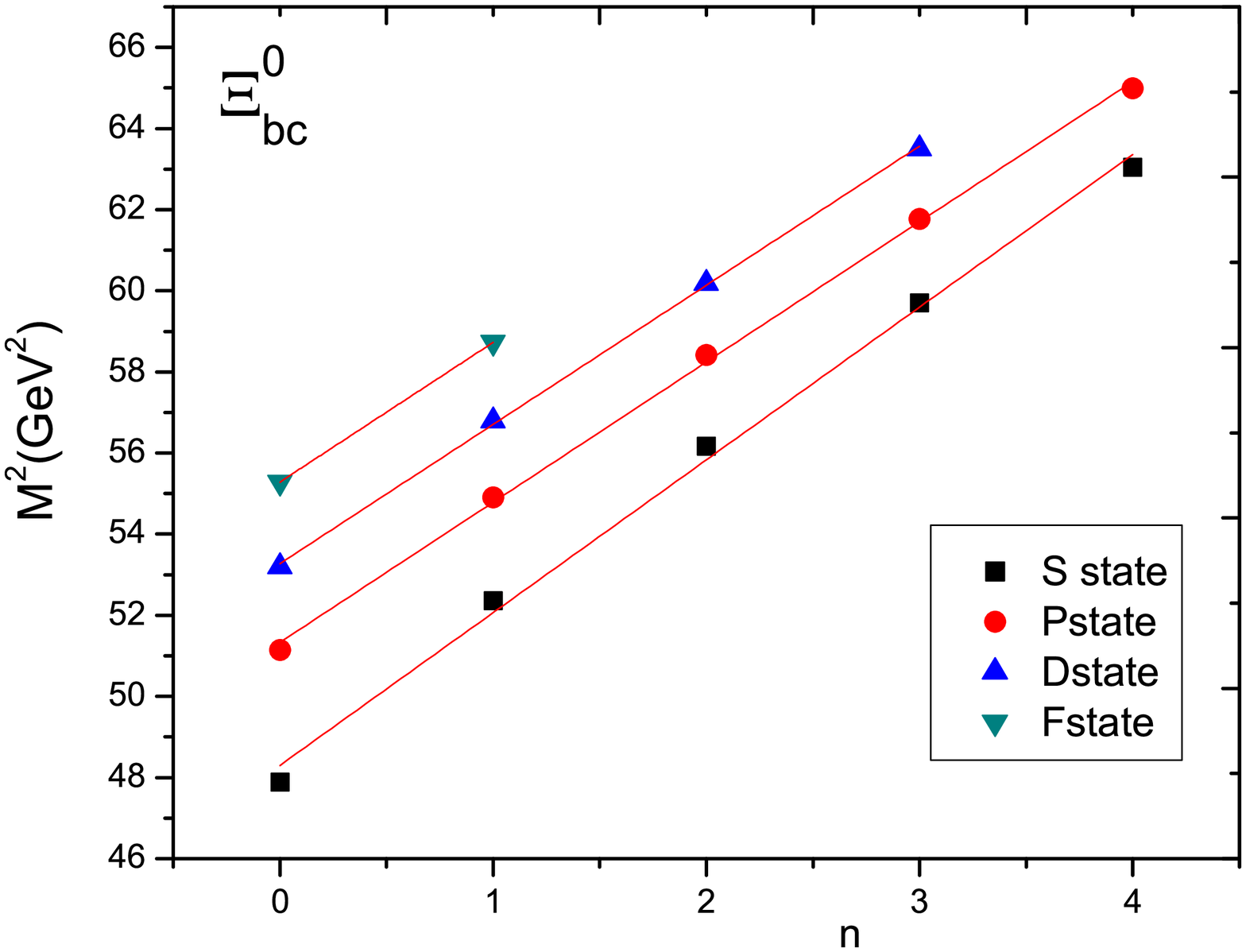}
}
\caption{Regge Trajectory ($M^{2}$ $\rightarrow$ n) for $\Xi_{bc}$ baryon.}
\label{fig:2}       
\end{figure}
 \begin{figure*}
 \includegraphics[width=0.52\textwidth]{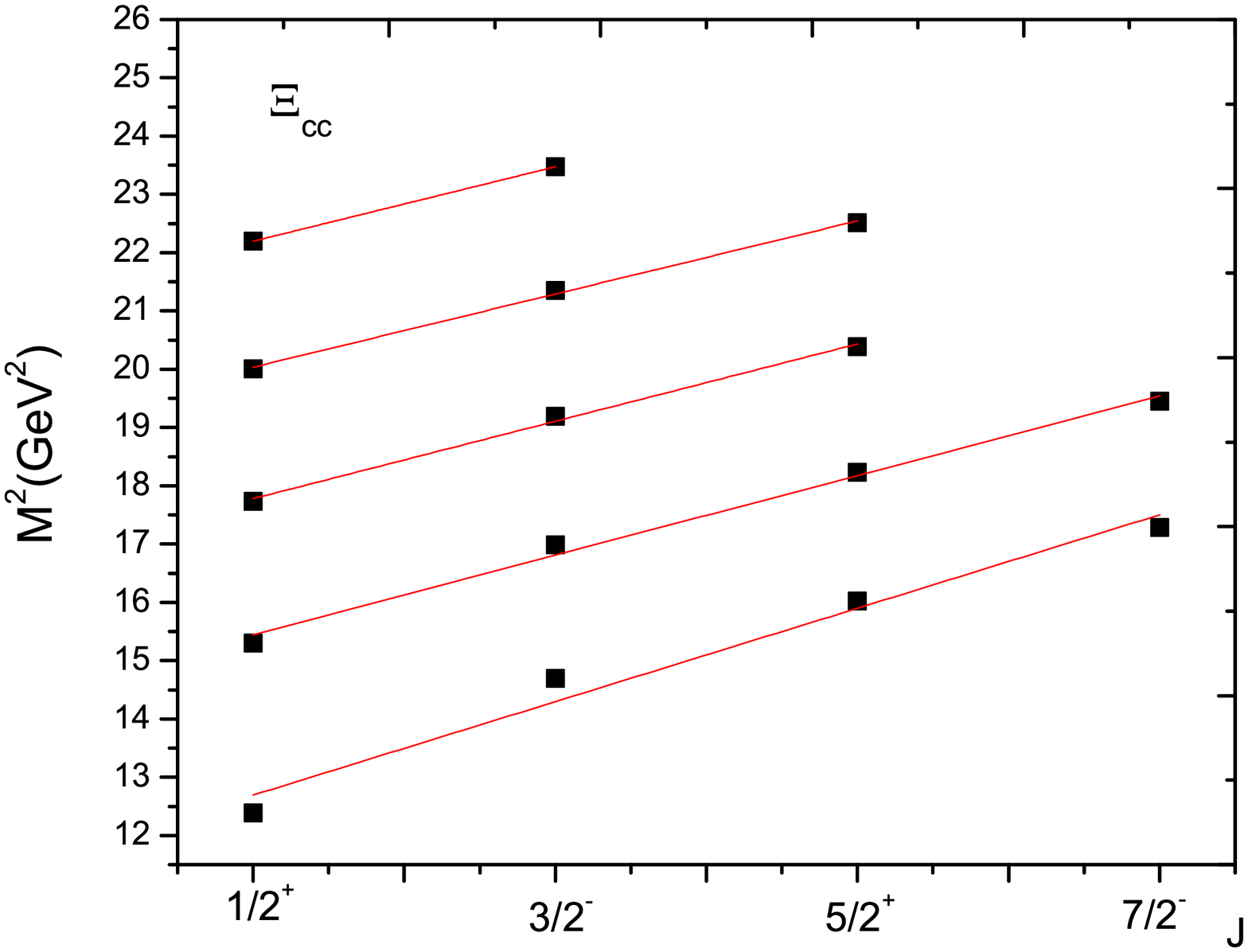}
  \includegraphics[width=0.52\textwidth]{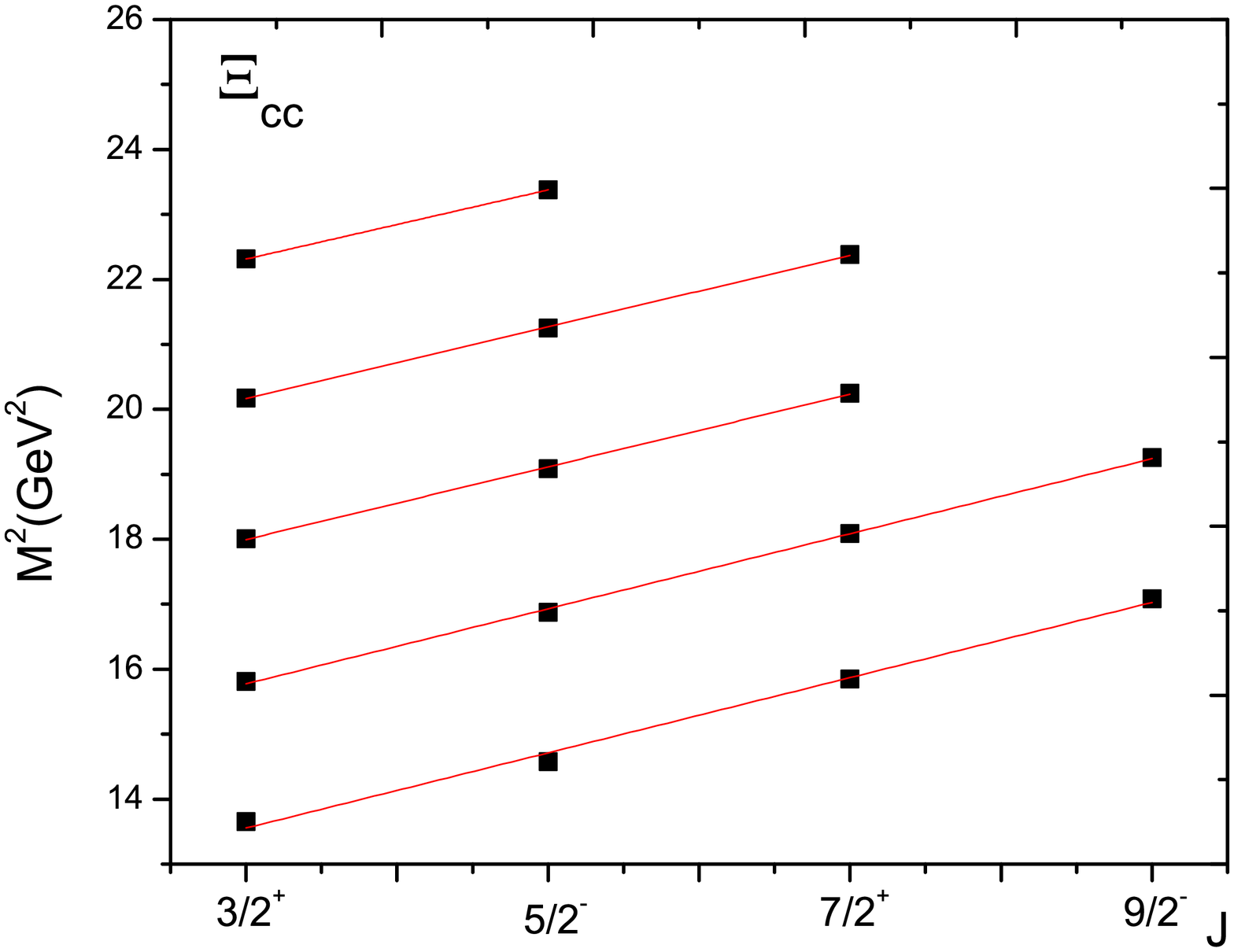}
  \caption{Regge Trajectory ($M^{2}$ $\rightarrow$ J) for $\Xi_{cc}$ baryon.}
\label{fig:3}  
\end{figure*}
 
\begin{figure*}
 \includegraphics[width=0.52\textwidth]{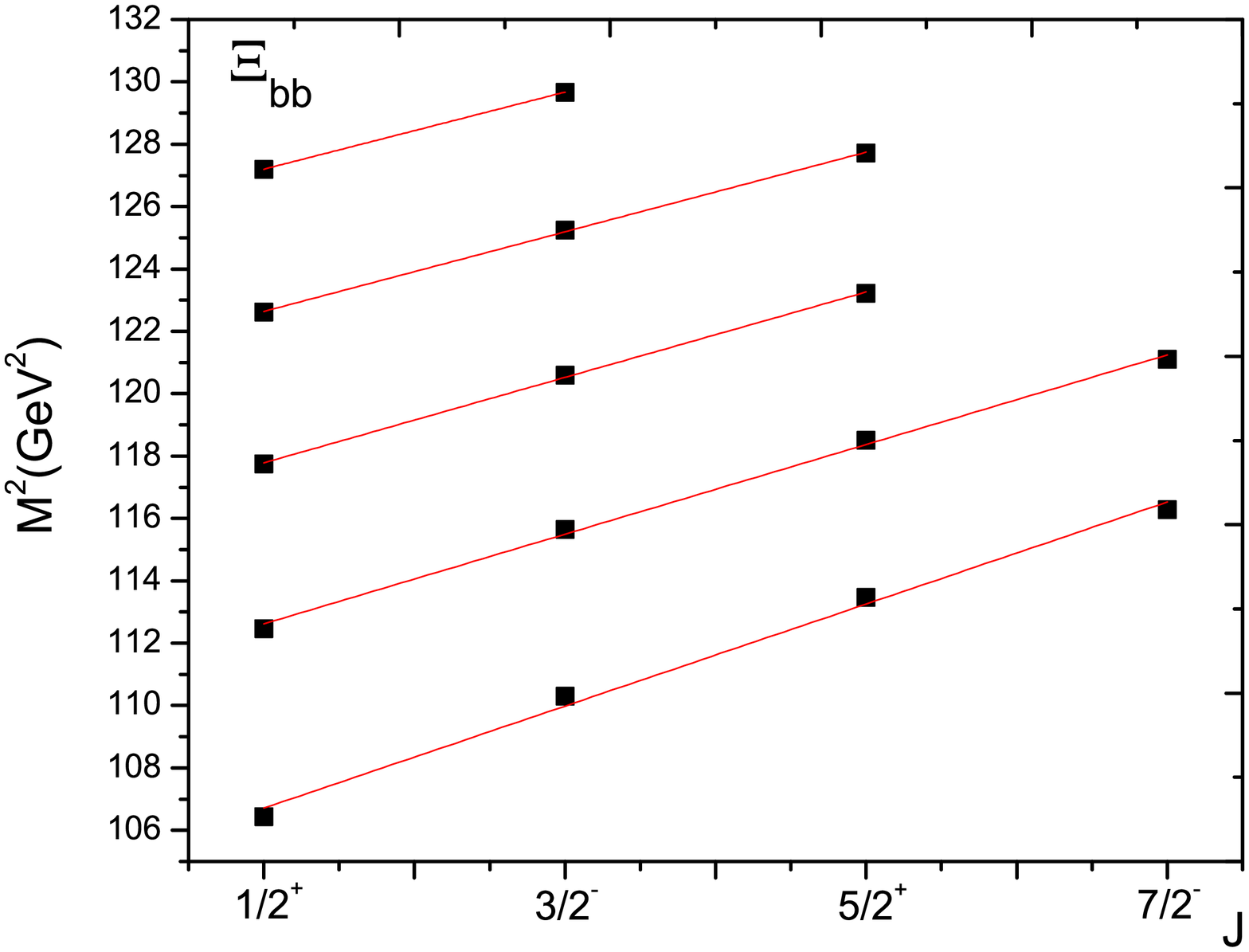}
  \includegraphics[width=0.52\textwidth]{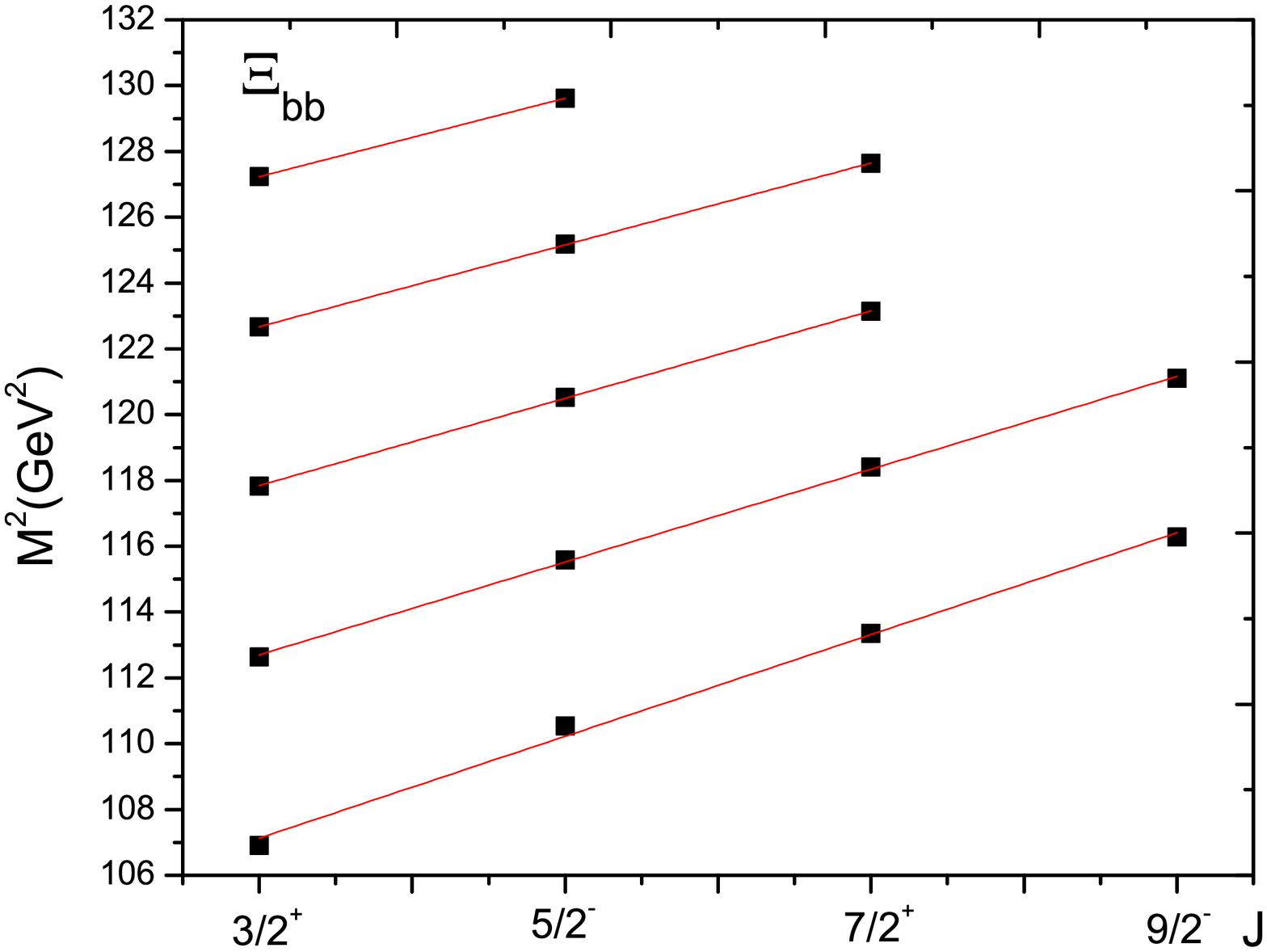}
\caption{Regge Trajectory ($M^{2}$ $\rightarrow$ J) for $\Xi_{bb}$ baryon.}
\label{fig:4}       
\end{figure*}
\begin{table*}
\centering
\caption{\label{table:8}Fitted parameters $\alpha$ and $\alpha_0$ are slope and Intercept of parent and daughter Regge trajectories.} 
\begin{tabular}{lllllll}
\hline\noalign{\smallskip}            
Baryon&Trajectory& $\alpha$&$\alpha_{0}$&$\alpha$&$\alpha_{0}$ \\
\hline
$\Xi_{ccd}$&parent&0.609$\pm$0.0607&-6.694$\pm$1.022&0.859$\pm$0.042&-10.650$\pm$0.648\\
&1 daughter&0.726$\pm$0.042&-10.197$\pm$0.737&0.865$\pm$0.016&-12.649$\pm$0.288\\
&2 daughter&0.751$\pm$0.041&-12.351$\pm$0.799&0.893$\pm$0.019&-15.069$\pm$0.368\\
&3 daughter&0.795$\pm$0.0319&-14.935$\pm$0.680&0.907$\pm$0.014&-17.292$\pm$0.298\\
\hline
$\Xi_{bbd}$&parent&0.304$\pm$0.0156&-31.433$\pm$1.733&0.322$\pm$0.013&-33.518$\pm$1.481\\
&1 daughter&0.347$\pm$0.0112&-38.094$\pm$1.312&0.354$\pm$0.005&-38.947$\pm$0.561\\
&2 daughter&0.365$\pm$0.007&-41.989$\pm$0.945&0.377$\pm$0.003&-43.419$\pm$0.344\\
&3 daughter&0.391$\pm$0.006&-46.983$\pm$0.772&0.402$\pm$0.003&-48.306$\pm$0.326\\
\noalign{\smallskip}\hline  
\end{tabular}
\end{table*}
\begin{table*}
\centering
\caption{\label{tab:table10} Magnetic Moment(in nuclear magnetons) of $J^{P}$~ $\frac{1}{2}^{+}$ and $\frac{3}{2}^{+}$ doubly heavy baryons.}
\begin{tabular}{llllllllll}
\hline\noalign{\smallskip}
Baryons& Magnetic moment&Our&\cite{patel}&\cite{103}&\cite{104}&\cite{101}\\
\hline
  $\Xi_{cc}^{+}$& $\frac{4}{3}\mu_{c}$- $\frac{1}{3}\mu_{d}$&0.784&0.859&0.722&0.80&0.785\\
  $\Xi_{cc}^{++}$& $\frac{4}{3}\mu_{c}$- $\frac{1}{3}\mu_{u}$&0.031&-0.133&0.114&-0.12&-0.208\\
   $\Xi_{bb}^{-}$&$\frac{4}{3}\mu_{b}$- $\frac{1}{3}\mu_{d}$&0.196&0.190&0.086&0.215&0.251\\
   $\Xi_{bb}^{0}$&$\frac{4}{3}\mu_{b}$- $\frac{1}{3}\mu_{u}$&-0.663&-0.656&-0.432&-0.630&-0.742\\
   $\Xi_{bc}^{0}$&$\frac{2}{3}\mu_{b}$+$\frac{2}{3}\mu_{c}$-$\frac{1}{3}\mu_{d}$&0.354&0.476&0.068&0.480&0.518\\
    $\Xi_{bc}^{+}$&$\frac{2}{3}\mu_{b}$+$\frac{2}{3}\mu_{c}$-$\frac{1}{3}\mu_{u}$&-0.204&-0.400&-0.236&-0.369&-0.475\\
 $\Xi_{cc}^{+ *}$& $2\mu_{c}$+$\mu_{d}$&0.068&-0.168&0.163&0.035&-0.311\\
 $\Xi_{cc}^{++ *}$& $2\mu_{c}$+$\mu_{u}$&2.218&2.749&2.001&-&2.670\\
$\Xi_{bb}^{- *}$&2$\mu_{b}$+$\mu_{d}$&-1.737&-0.951&-0.652&-1.029&-0.522\\
$\Xi_{bb}^{0 *}$&2$\mu_{b}$+$\mu_{u}$&-1.607&1.576&0.916&1.507&1.870\\
$\Xi_{bc}^{0 *}$&$\mu_{b}$+$\mu_{c}$+$\mu_{d}$&-0.372&-0.567&-0.257&-0.508&-0.712\\
$\Xi_{bc}^{+ *}$&$\mu_{b}$+$\mu_{c}$+$\mu_{u}$&1.562&2.052&1.414&2.022&2.270\\
\noalign{\smallskip}\hline
\end{tabular}
\end{table*}

\begin{equation}
J=\alpha M^2+ \alpha_{0}
\end{equation}
\noindent Where, $\alpha$ and $\alpha_{0}$ are slope and intercept, respectively. The fitted slopes and intercepts are given in Table~\ref{table:8}. 
We observe that the square of the calculated masses fit very well to the linear trajectory and almost parallel and equidistant in S, P, D and F states. We can determine the possible quantum numbers and prescribe them to particular Regge trajectory with the help of our obtained results.\\

To obtain magnetic moments of the $\Xi$ family, we need to calculate their effective masses first. As the combination of quarks in baryons changes, its binding interaction affects and $m_{i}^{eff}$ differs. The effective mass for each of the constituting quark $m_{i}^{eff}$ can be defined as
\begin{equation}
m_{i}^{eff}= m_i\left( 1+ \frac{\langle H \rangle}{\sum_{i} m_i} \right)
\end{equation}
where, $\langle H \rangle$ = E + $\langle V_{spin} \rangle$. Thus, the magnetic moment of baryons with bound quarks are given as [\cite{zalak1} and Ref. their in]
\begin{eqnarray}
\mu_{B}=\sum_{i}\langle \phi_{sf}\vert \mu_{iz}\vert\phi_{sf}\rangle)
\end{eqnarray}
where
\begin{equation}
\mu_{i}=\frac{e_i \sigma_i}{2m_{i}^{eff}}
\end{equation}
$e_i$ is a charge and $\sigma_i$ is the spin of the respective constituent quark corresponds to the spin flavor wave function of the baryonic state. Using these equations and our obtained ground state masses(mentioned in Table ~\ref{table:1}), we calculated magnetic moments of all six $\Xi$ baryons. The spin flavor wave function and magnetic moments are given in Table {~\ref{tab:table10}. Our obtained ground state magnetic moments are also compared with others shown in Table {~\ref{tab:table10}.

\section{Conclusion}

The Hypercentral constituent quark model is used to construct the mass spectra of doubly heavy $\Xi$ baryons. Ground states as well as excited state masses are obtained successfully. Mass difference between the light quarks (u and d) is 12 MeV in our model. So, it is obvious that when we move towards the calculations of the excited states the baryon masses would also have a very small mass difference. For the sake of completeness we calculated whole mass spectrum for all six doubly heavy baryon and noticed that it hardly differs less than $\approx$10 MeV; which can be observed in Tables ~\ref{table:1}- \ref{table:6}. The ground state of $\Xi_{cc}$ is experimental known and while comparing  our ground states of $\Xi_{cc}^{++}$,$\Xi_{cc}^{+}$ baryons we define the state with parity $J^{P}$= $\frac{1}{2}^{+}$. 

We successfully plotted Regge trajectories of present work in both (n, $M^{2}$) and (J,$M^{2}$) planes and fortunately assigned the quantum number for each cases of six $\Xi$ baryons. The magnetic moments of the ground states are also calculated using obtained masses. We can observe that our obtained results are close to other predictions(except $\Xi_{cc}^{++}$,$\Xi_{cc}^{+ *}$, $\Xi_{bb}^{0 *}$ baryons).

This study will definitely help future experiments and other theoretical models to identify the baryonic states from resonances. We would like to extend this model to calculate the mass spectra and other properties of triply heavy baryons.
\begin{acknowledgements}
A. K. Rai acknowledges the financial support extended by DST, India under SERB fast track scheme SR/FTP/PS-152/2012. We are very much thankful to Prof. P. C. Vinodkumar and Dr. Kaushal Thakkar for their valuable suggestions throughout the work.
\end{acknowledgements}


\begin{thebibliography}{}

\bibitem{zalak1} Z. Shah, K. Thakkar and  A. K. Rai, Eur. Phys. J. C \textbf{76}, 530 (2016).
\bibitem{olive}C. Patrignani et al.[Particle data Group], Chin. Phys. C \textbf{40}, 100001 (2016). 
\bibitem{selex} M. Mattson et al.[SELEX Collaboration], Phys. Rev. Lett. \textbf{89}, 112001 (2002).  
\bibitem{selex1}  A. Ocherashvili et al. [SELEX Collaboration], Phys. Lett. B \textbf{628}, 18 (2005).
\bibitem{babar}B. Aubert et al. [BABAR Collaboration], Phys. Rev. D \textbf{74}, 011103 (2006).
\bibitem{belle} R. Chistov et al. [BELLE Collaboration], Phys. Rev. Lett. \textbf{97}, 162001 (2006).
\bibitem{lhcb}R. Aaij et al. [LHCb Collaboration], J. High Energy Phys. \textbf{12}, 090 (2013).
\bibitem{focus}S. P. Ratti, Nucl. Phys. Proc. Suppl. {\bf 115}, 33 (2003).
\bibitem{yoshida}  T. Yoshida, E. Hiyama, A. Hosaka, M. Oka, and K. Sadato, Phys. Rev. D \textbf{92}, 114029 (2015)
\bibitem{kwei} K. W. Wei, B. Chen and  X. H. Guo, Phys. Rev. D \textbf{92} 076008 (2015);  K. W. Wei et al. {\it arXiv:1609.02512v1 [hep-ph]} (2016)
\bibitem{paula} P. P. Rubio, S. Collins, and G. S. Baliy, Phys. Rev. D \textbf{92}, 034504 (2015)
\bibitem{brown} Z. S. Brown, W. Detmold, S. Meinel, and K. Orginos, Phys. Rev. D \textbf{90}, 094507 (2014)
\bibitem{alex} C. Alexandrou, V. Drach, K. Jansen, C. Kallidonis, and G. Koutsou, Phys. Rev. D \textbf{90}, 074501 (2014)
\bibitem{mathur} M. Padmanath, R. G. Edwards, N. Mathur, and M. Peardon,  Phys. Rev. D {\bf 91}, 094502 (2015) 
\bibitem{aliev}  T. M. Aliev, K. Azizi, and M. Savci, Nucl. Phys. A \textbf{895}, 59 (2012); J. Phys. G \textbf{40},  065003 (2013)
\bibitem{roberts}  B. Eakins and W. Roberts, Int. J. Mod. Phys. A \textbf{27}, 1250039 (2012)
\bibitem{Roberts2008} W. Roberts and M. Pervin, Int. J. Mod. Phys. A \textbf{23}, 2817 (2008)
\bibitem{Giannuzzi2009} F. Giannuzzi, Phys. Rev. D \textbf{79}, 094002 (2009).
\bibitem{valcarce} A. Valcarce, H. Garcilazo, and J. Vijande, Eur. Phys. J. A \textbf{37}, 217 (2008) 
\bibitem{ebert} D. Ebert, R. N. Faustov, V. O. Galkin and A. P. Martynenko, Phys. Rev. D \textbf{66}, 014008 (2002)
\bibitem{Gershtein} S.S. Gershtein, V.V. Kiselev, A.K. Likhoded, A.I. Onishchenko, Phys. Rev. D 62, 054021 (2000).
\bibitem{kar} M. Karliner and J. L. Rosner, Phys. Rev. D {\bf 90}, 094007 (2014)
\bibitem{Ltang}L. Tang, X-H Yuan, C-F Qiao, and X-Q Li, Commun. Theor. Phys. \textbf{57}, 435 (2012)
\bibitem{albertus}C. Albertus, E. Hernandez, J. Nieves1, and J.M. Verde-Velasco, Eur. Phys. J. A \textbf{32}, 183 (2007). 
\bibitem{1}Martynenko A P 2008 {\it Phys. Lett.} B \textbf{663} 317 
\bibitem{wang} Z. G. Wang, Eur. Phys. J. A  \textbf{47}, 267 (2010); Eur. Phys. J. C \textbf{68}, 459 (2010)
\bibitem{patel}B. Patel, A. K. Rai and P. C. Vinodkumar, Pramana J. Phys. \textbf{70}, 797 (2008); J. Phys. G \textbf{35},  065001 (2008); Phys. Rev. C \textbf{78}, 055202 (2008)
\bibitem{zalak} Z. Shah, K. Thakkar, A. K. Rai and P.C. Vinodkumar, Eur. Phys. J A \textbf{52}, 513 (2016)
\bibitem{zalak3}  Z. Shah, K. Thakkar, A. K. Rai and P.C. Vinodkumar,  Chin. Phys. C \textbf{40}, 123102 (2016).
\bibitem{Bijker}R Bijker, F Iachello and A Leviatan, Ann. Phys. \textbf{284}, 89 (2000)
\bibitem{ginnani2015}E. Santopinto, F. Iachello, M.M. Giannini, Eur. Phys. J. A 1, 307-315 (1998)
\bibitem{M. Ferraris}M. Ferraris, M. M. Giannini, M. Pizzo, E. Santopinto and L. Tiator, Phys. Lett. B \textbf{364}, 231 (1995)
\bibitem{koma} Y. Koma, M. Koma, H. Wittig, Phys. Rev. Lett \textbf{97}, 122003 (2006)
\bibitem{11}N. Devlani, V. Kher, and A. K. Rai, Eur. Phys. J. A \textbf{50}, 154 (2014); Eur. Phys. J. C \textbf{75}, 462
(2015)
\bibitem{12}M. B. Voloshin, Prog. Part. Nucl. Phys. \textbf{51}, 455 (2008)
\bibitem{lucha} W. Lucha and F. Schoberls, Int. J. Mod. Phys.C. {\bf 10}, 607 (1999)
\bibitem{103} A. Bernotas and V. Simonis, {\it arXiv:1209.2900v1} (2012)
\bibitem{104}R. Dhir and R. C. Verma, Eur. Phys. J. A \textbf{42}, 243 (2009); Phys. Rev. D {\bf 88}, 094002 (2013)
\bibitem{101}C. Albertus, E. Hernandez, J. Nieves and J. M. Verde-Velasco, Eur. Phys. J. A \textbf{32}, 183 (2007)




\end{thebibliography}
\end{document}